\documentclass[aip,twocolumn,letterpaper]{revtex4}

\usepackage{fullpage}

\usepackage{graphics}
\usepackage{epsfig}

\usepackage[margin=1.0in]{geometry}

\begin{document}
\title{Needed computations and computational capabilities for stellarators}
\author{Allen H. Boozer}
\affiliation{Columbia University, New York, NY  10027}

\begin{abstract} 
 Stellarator plasmas are externally controlled to a degree unparalleled by any other fusion concept, magnetic or inertial.  This control is largely through the magnetic fields produced by external coils.   The development of fusion energy could be expedited by carrying out remarkably straight-forward computations to define strategies for exploiting this external control.  In addition to these computations, which have a reliability limited only by competence, certain physics areas that affect the develop of stellarator power plants should have more intense study.  The low cost and speed with which computations can be carried out relative to experiments has implications for the development of fusion.  Computations should be used to develop a strategy that to the extent possible allows major issues to be circumvented.  Required computations for this strategy are the subject of this paper.

\end{abstract}

\date{\today} 

\maketitle
%\tableofcontents

\section{Introduction}

Stellarators uniquely allow computational design to circumvent major issues.  Many were discussed in 2015 in the paper \emph{Stellarator design} \cite{Boozer:design}.  The coils that define the magnetic configuration make stellarator plasmas externally controlled to an extent unmatched by any other fusion concept, magnetic or inertial.  The focus of stellarator design need not be extrapolation, with its conservatism and dangers from changing physics regimes, see Section 2 of \cite{Boozer:CO2-Stell}, but innovative solutions, which could speed the development of attractive fusion power. 

Curl-free magnetic fields are the natural starting point for studies of stellarator configurations.  They can be extended to stellarators with arbitrary pressure profiles by using a perturbed equilibrium solver such as the DESC code \cite{DESC:2020}.  By increasing the plasma pressure in a number of computational steps, the externally-produced normal magnetic fields can be adjusted to access different values and profiles of the plasma pressure.   This computational strategy models the natural startup of a stellarator fusion power plant.

Innovative computational design has low demands on financial resources but high demands on organizational skills.  Creative people are naturally drawn into programs in which their ideas could have global implications.  Important innovations should be rewarded by study and not diluted by studies of ideas that violate fundamental physics or engineering constraints.   Organizational skill requires knowing the difference and inspiring thought on what remaining issues in fusion energy could be circumvented by clever design.  Clever design is particularly important for the coils, which can be studied with a reliability limited only by competence and give the primary external control of toroidal fusion plasmas.

Appendix \ref{sec:B-freedom} gives the limitations that mathematics and Maxwell's equations place on the freedom of the use of external coils for the control of a toroidal plasma.  The general external magnetic field in the plasma region can be written in $(R,\varphi,Z)$ cylindrical coordinates as
\begin{equation}
\vec{B}_x= \frac{\mu_0G_0}{2\pi R}\hat{\varphi} + \vec{\nabla}\phi. \label{gen-B}
\end{equation}
$G_0$ is the number of Amperes of poloidal current, which comes up through the hole in the torus, and $\phi$ can be given by a single-valued solution to Laplace's equation $\nabla^2\phi=0$.  A Neumann boundary condition, $\hat{n}\cdot\vec{\nabla}\phi$, on any surface that encloses the toroidal plasma but excludes the coils, specifies $\phi$ uniquely except for an irrelevant additive constant. Equivalently,  $\phi$ can be represented by magnetic dipoles oriented normal to the enclosing surface.   In the limit of a dense coverage of the enclosing surface by dipoles, the dipoles represent a single-valued current potential, Appendix \ref{sec:B-freedom}.  Localized currents can be represented by patches of dipoles, which are a discrete representation of a localized current potential, called a current potential patch by Todd Elder \cite{Elder-Boozer:2024}.

A net toroidal current flowing outside the enclosing surface, which can be produced by poloidal-field or helical coils, controls the poloidal magnetic flux through the hole in the torus.  Changes in the poloidal flux in the hole of the torus produces a loop voltage in the region interior to the enclosing surface, but the magnetic field due to a net toroidal current can be represented in the interior region by dipoles on the enclosing surface.

Two features of optimized stellarators, which appear to be important for circumventing coil issues, have had little exploitation.  The first is a result of Ku and Boozer  in 2010 and 2011 \cite{Ku-Boozer:2010,Boozer-Ku:control2011}.  Only a small fraction of the possible linear external perturbations $\delta\vec{B}_x\cdot\hat{n}$ at the surface of an optimized stellarator equilibria have an effect on the degree of physics optimization.  The ones that do, such as the ones that affect global neoclassical transport, can be organized into external field distributions  that decrease exponentially in importance from one distribution to the next.  The second is a result of Elder and Boozer \cite{Elder-Boozer:2024}.   The general field of Equation (\ref{gen-B}) with dipoles covering only $\sim 25\%$ of an enclosing surface at the location of the HSX stellarator coils can accurately reproduce their field on the plasma.

%%%%%%%%%%%%%%%%%%%%%%%

\section{Fundamental coil choices \label{sec: coil choices}}

Since stellarators are intrinsically helical, helical coils are a natural solution.  The most famous example is LHD \cite{LHD:1996}.  By definition helical coils encircle the plasma both toroidally and poloidally, so they can produce the required toroidal flux as well as shape the plasma.  As shown by Yamaguchi \cite{Yamaguchi}, helical coils have a number of advantages.  They have relatively open access to the plasma, do not have the coil ripple associated with toroidal field coils, and can be consistent with good particle confinement.   Despite the success in winding the superconducting helical coils in LHD, the difficulty of winding such coils is considered prohibitive in a power plant---that is unless a way can be found to assemble a superconducting coil in segments.  

In 1972 Wobig and Rehker \cite{Modulars:1972} published a discovery that revolutionized  stellarator coil design.  The effect of helical coils can be produced by modular coils.  Modular coils, which are shaped toroidal field coils, can make both the toroidal and the helical fields required for a stellarator.  Their much smaller size compared to helical coils simplifies the construction of stellarators.  

In 1987, Merkel \cite{Merkel:1987} published the code NESCOIL, which calculates the modular coils lying outside an enclosing toroidal surface that are needed to support a given plasma equilibrium.  The FOCUS code \cite{Focus code: 2017}, which was published in 2017, finds modular coils that satisfy given physics and engineering constraints using closed space curves.

Nevertheless, modular coils have issues. They restrict plasma-chamber access, have intrinsic ripple, and provide an unnecessary restriction on plasma-coil separation.  The plasma-coil separation issue can be understood in a straight-cylindrical $(r,\theta,z)$  model.  The toroidal, or $\hat{z}$-directed, field and the number of Amperes required to produce it are independent of $r$.  Fields that depend on the poloidal angle $\theta$, as in $\sin m \theta$, increase as $r^{m-1}$ as do the magnitude of currents required to produce them.  Such fields are required in a stellarator: $m=2$ for ellipticity, $m=3$ for triangularity, etc.  The implication is that modular coils for stellarators become more convoluted in shape the further back they are from the plasma.  Current potential patches may allow a larger plasma-coil separation than can modular coils alone.  The importance of this separation is discussed in Section \ref{Sec:separation}, and the issue was studied in Todd Elder's thesis \cite{Elder:thesis2023}. 

Although stellarators designed for a path to fusion generally use modular coils, Thea Energy \cite{Thea Energy:2023} presented a coil design in 2023 that consists of hundreds of planar coils.  A few of these coils are toroidal field coils, but most are identical small coils, which are individually controllable and allow the specification of the normal component of the magnetic field over an enclosing surface.  

The independent use of a large number of coils to control plasma configurations has a fundamental mathematical limitation related to the condition number of matrices.  This limitation is explained in Section \ref{Sec:Control Matrix} and can and should be studied computationally.

The magnetic fields associated with current-potential patches on an enclosing surface require coils be engineered.  Attractive compact coils presumably exist when the magnetic field that a current-potential patch produces has a field strength that is highly constrained in its magnitude and its spatial gradients.  %Attractive designs for compact coil sets  \cite{Boozer:2009}, Figure (\ref{fig:removable-coils}), that can be easily removed along with sections of the enclosing wall, whenever plasma chamber access is required, can and should be computationally designed.

%%%%%%%%%%%%%%%%%%%%%%%%

\subsection{Coils allowing easy plasma chamber access \label{sec: access}}

There is no requirement that the plasma-encircling coils, such as modular coils, be chosen to ensure $\vec{B}\cdot\hat{n}=0$ on the plasma surface.  But, they should be chosen to ensure good plasma access, a large separation from the plasma, Section \ref{Sec:separation}, and may be chosen using other criteria such as simplicity.  Construction difficulties associated with maintaining high spatial precision in the coils can be greatly reduced and will be discussed in Section \ref{Sec:precision}.

Access affects both the cost of fusion power and the speed with which fusion systems can be developed.  

The 14~MeV neutrons from deuterium-tritium (DT) fusion reactions degrade materials unacceptably after a certain fluence, which is approximately 10~MW yr/m$^2$.  Fusion energy is cheaper the greater the power density, MW/m$^2$ through the walls until material limits are reached.  A major limitation on the fraction of time a fusion power plant can actually produce power, its duty factor, is the speed with which neutron-degraded plasma-chamber components can be replaced.  Even in ITER \cite{ITER hot Cell: 2017}, the replacement of degraded components is complicated by both their high level of radioactivity and by the requirement of maintaining confinement of tritium.  The required replacement time is minimized by open plasma-chamber access.  

Open access is not only important for rapid component replacement in power plants but also for testing different designs for in-vessel components during fusion development.  These components are a major challenge to fusion.

%%%%%%%%%%%%%%%%%%%%%%
\begin{figure}
\centerline{ \includegraphics[width=2.5 in]{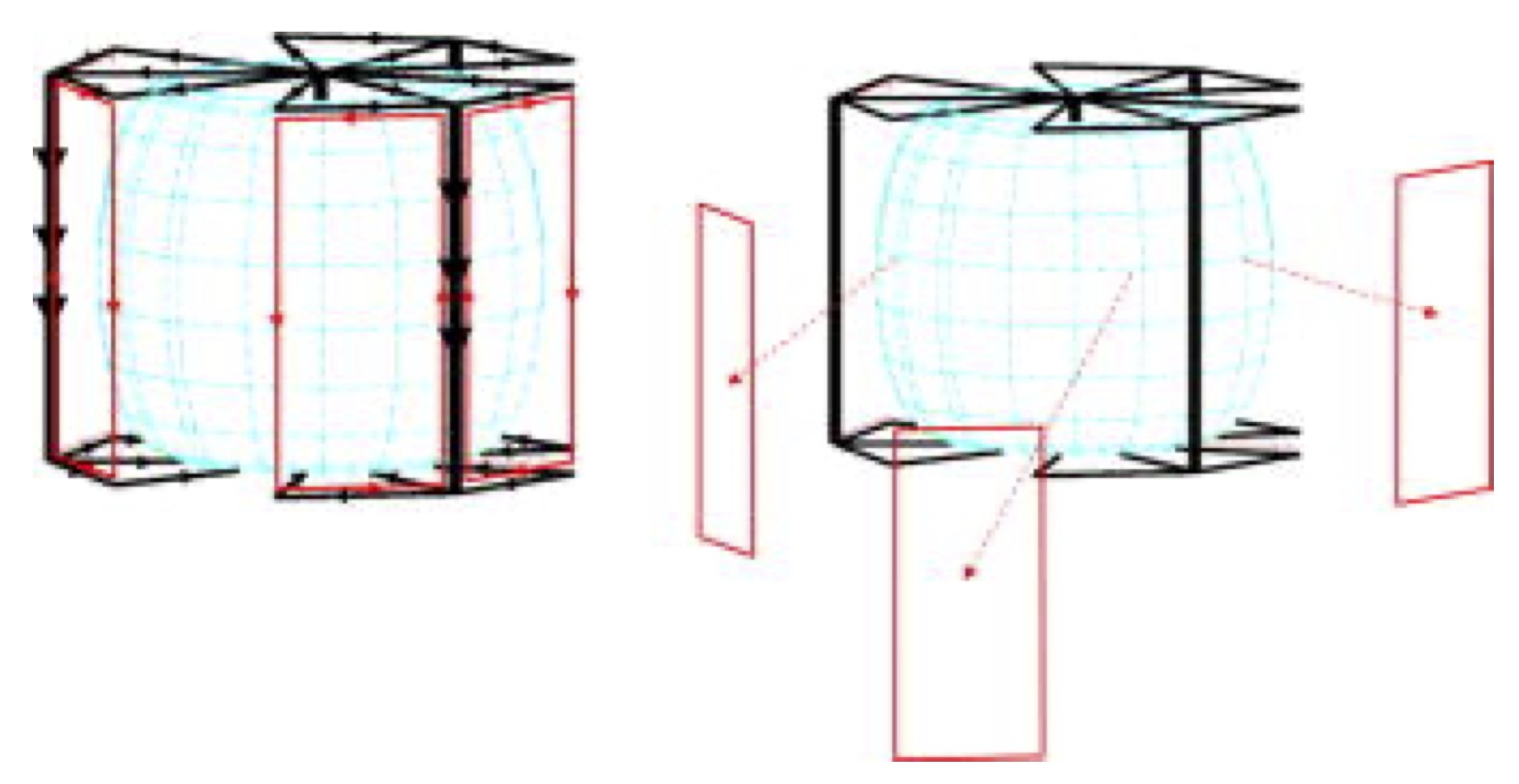}}
\caption{Non-plasma encircling coils mounted on removable wall sections can be used to provide open access to fusion plasmas.  This was illustrated in Figure (4) of \cite{Boozer:2009} in  2009 as a schematic way to enhance plasma access in tokamaks.} 
\label{fig:removable-coils}
\end{figure}
%%%%%%%%%%%%  

Two concepts for open access to the plasma chamber are:  (1) large spaces between coils and (2) non-plasma encircling coils that can be removed along with a large section of the chamber wall \cite{Boozer:2009}, Figure (\ref{fig:removable-coils}).  

Large spaces between the coils are possible with helical coils \cite{Yamaguchi} and are advantageous for heating and diagnostic ports as well as for in-vessel changes.  Ripple in the magnetic field strength places difficult constraints on the space between modular coils unless mitigated by  coils mounted on removable wall sections.  

The use of non-plasma encircling coils mounted on removable wall sections have few fundamental constraints.  Coils that can be removed along with the part of the vessel with which they are associated can be represented by current potential patches, which are patches of magnetic dipoles \cite{Elder-Boozer:2024}.  When a section of the vessel is removed for plasma-chamber access, it must be placed back into position with a high accuracy to obtain a vacuum seal---probably with more accuracy than is required for the placement of the coils.

Computational studies of what is and what is not possible in coil designs with large removable wall sections are needed.  Such studies are themselves clearly feasible,  and would be a basis for reliable designs for stellarators with open maintenance-access to the plasma chamber. 

%%%%%%%%%%%%%%%%%%%%%%%%%%%%%%%%%%%%%%%%%%%%%%%

\subsection{Coils with maximal separation from the plasma \label{Sec:separation}}

The thickness of tritium-breeding blankets and neutron shields sets a fundamental limit on the separation between the plasma and the coils.   The traditional estimate is approximately 1.5~m.  Whatever this scale may be, it sets a minimum spatial scale for a deuterium-tritium (DT) fusion power plant. The larger the coil-plasma separation that can be consistent with feasible coils,  the more freedom is available for designing DT power plants having a smaller power and cost per plant.

Maximization of coil-plasma separation requires a simultaneous optimization of three quantities in the choice of coils.  

The first is a measure of the desirable physics properties of the magnetic configuration, such as acceptable neoclassical transport.  This can be defined by the shape of a reference surface, an outer magnetic surface. 

The second is a measure of how efficiently distant coils can support a magnetic surface of that shape.  For curl-free magnetic configurations, external coils must produce a field normal to the reference surface  $b(\theta,\varphi)=-(\mu_0G_0/2\pi R)\hat{\varphi}\cdot\hat{n}$, where $\hat{n}$ is the unit normal to the reference surface.  Coil efficiency drops exponentially with separation \cite{Boozer:RMP}.  The coefficient in the exponential is approximately given by the wavenumber of the variation of $b$ in the reference surface \cite{Landreman:field-gradient}.  This result follows from a two-dimensional model; $\nabla^2\phi=0$ implies $\phi$ has the typical dependence $\sin(k_y y) \exp(-k_x x)$ with $k_x=k_y$.  The exponential drop in efficiency implies that crude measures, such as the strength of the dipoles required to produce the field, $\vec{d}^\dag\cdot\vec{d}$ of Appendix \ref{sec:B-freedom}, provide a useful measure.  

The third is a measure of how the efficiency of the coils represented by dipoles is reduced by enforcing open access requirements. Chamber access and coil-plasma separation interact in determining where the patches of dipoles should be located.

%%%%%%%%%%%%%%%%%%%%%%%%%%%%%%%%%%%%%%%%%%%%%%%%%%%%
%%%%%%%%%%%%%%%%%%%%%%%%%%%%%%%%%%%%%%%%%%%%%%%

\subsection{Coils allowing flexibility \label{sec:flexibility} }

There are two types of external perturbations: (1) ones that are consistent with the periodicity of the optimized configuration and (2) those that are not.  The first type are considered changes in the configuration while the second type are considered to be error fields, the effect of construction inaccuracies.  The orders of magnitude differences in configuration sensitivity to changes in different external magnetic field distributions \cite{Ku-Boozer:2010,Boozer-Ku:control2011} should be recognized in stellarator coil design.   Designing coils by minimizing $\vec{B}\cdot\hat{n}$ on a given surface ignores these differences.

Any external magnetic-field distribution to which the plasma is sensitive should be controllable by adjustments to the coil currents  \cite{Boozer:design}.   These field distributions could be determined from the Hessian matrix of a plasma optimization. 

 Unfortunately, there is a mathematical limit on the number of currents that can be adjusted with sufficient accuracy to control physics properties.  This limit is approximately given by the condition number of a matrix, Section \ref{Sec:Control Matrix}.

%%%%%%%%%%%%%%%%%%%%%%%%%%%%%%%%%%%%%%%%%%%%%%%

\subsection{Coils that reduce precision requirements \label{Sec:precision}}

The orders of magnitude difference in sensitivity to different error-field distributions implies coil systems can be easily designed to greatly reduce the required accuracy of construction.  This was explained at a 2010 ITER workshop \cite{Boozer:tok-error2011} but ignored in the design of the ITER error-field correction coils.  

The calculation of control or error field sensitivities has been largely ignored in stellarator-design studies as well as in tokamak studies.  Nevertheless, it is difficult to grasp how a coil system can be considered to be optimized without such a study.

Once a desirable stellarator plasma has been found, a smooth surface can be defined that is on the plasma side of the coils but is otherwise as far from the plasma as possible.  Consider the effect of an arbitrary Fourier decomposed current potential on this surface, $\kappa(\theta,\varphi)$.  Using a stellarator optimization code, the collection of Fourier amplitudes, with $\oint \kappa^2 d\theta d\varphi$ fixed, can be determined that maximally degrades the optimization.  This defines current potential, $\kappa_1$, that gives the worst possible external field error and frequently has an $n=1$ toroidal Fourier dependence.  The second worst external field error can be defined by determining the $\kappa_2$ with $\oint \kappa_2\kappa_1 d\theta d\varphi=0$ by the same method as $\kappa_1$.  That is, all the dangerous field errors can be determined in a series ordered by the plasma sensitivity to them.

There are two strategies for reducing the precision requirements of coil construction.  First, the plasma optimization can be used to reduce the plasma sensitivity to the worst magnetic-field error, which is defined by $\kappa_1$.  This efficiently accomplishes what is accomplished by a stochastic optimization  \cite{Stochastic optimization}. 

Secondly, error-field correction coils can be installed that can independently drive the various magnetic field distributions to which the plasma is sensitive.  These coils can make the total external field in distributions to which the plasma is sensitive zero.  External field errors are not cancelled.  The external drive just makes the externally-produced magnetic field distributions to which the plasma is sensitive zero.  Field errors to which the plasma is not sensitive are driven in this process, so there is maximum number of error field distributions that can be controlled in this way.  Fortunately, the plasma sensitivity appears to drop exponentially, so only a few error field distributions need to be controlled to greatly reduce the precision requirements on the coils.

%%%%%%%%%%%%%%%%%%%%%%%%%%

\subsection{Coil designs consistent with major configuration changes \label{sec: changes}}

As discussed in Section \ref{sec:flexibility}, plasma sensitivity to small changes in the external magnetic field defines external field distributions that should be controllable in order to have experimental flexibility.  However, far more radical changes of the plasma configuration are possible---for example changing the plasma shape from a quasi-isodynamic to a quasi-helically symmetric equilibrium.  A fundamental limit on radical changes is set by the plasma not intercepting non-replaceable parts of the chamber walls.  More subtle limits are set by the locations of divertors and diagnostics.

Todd Elder's result that approximately 25\% coverage of an enclosing surface by current potential patches allows the accurate definition of a quasi-helically symmetric equilibrium suggests two strategies:  (1) Define current-potential patches that are consistent with major changes in the configuration.  (2) When the current-potential  patches are on removable sections of the chamber walls, one set of wall inserts carrying a current-potential patch could be replaced by another.  The first option sounds preferable, but changes in the surrounding walls may be sufficiently great that the second option is required.

%%%%%%%%%%%%%%%%%%%%%%%%%%%%%%%%%%%%%%%%%%%%%%%

\section{Control Matrix Issues \label{Sec:Control Matrix}}

The control matrix $\tensor{C}$ relates a matrix vector $\vec{I}$ that has elements equal to the currents in the control coils to the parameters that one one wishes to control, the important perturbations to the external magnetic field  $\delta\vec{B}_x\cdot\hat{n}$ on the plasma surface.  That is 
\begin{equation}
\vec{P}=\tensor{C}\cdot \vec{I}.
\end{equation}   
The external field perturbations on the plasma surface are of two types: (1) ones that are consistent with the periodicity of the optimized configuration, Section \ref{sec:flexibility}, and (2) those that are not---the error fields, Section \ref{Sec:precision}.

The mathematics of the two types of perturbations is somewhat different.  The first type is defined by external magnetic fields $\delta\vec{B}_x\cdot\hat{n}$ associated with large terms in the Hessian matrix.  The second type is defined by the current potentials $\kappa_1$, $\kappa_2, \cdots$ to which the plasma is sensitive.  For both types, the current potentials that would produce the precise perturbation would be be non-zero over a large fraction of the surface on which they are defined and are generally sensitive to the plasma state that one is trying to control.

 To obtain practical coils, only patches of current potential can be retained, which should be in the places where the current potentials required for control are largest.  Although patches of current potential can control the quantities that are to be controlled, they do not allow the independent control of all possible $\delta\vec{B}_x\cdot\hat{n}$ distributions.  Compact current-potential patches produce crosstalk between the drives of the various $\delta\vec{B}_x\cdot\hat{n}$ distributions.  
 
 There is a subtle mathematical problem.  How is the choice of control coils to be made in order to have a control matrix $\tensor{C}$ that optimizes the control of the most important  $\delta\vec{B}_x\cdot\hat{n}$ distributions over the broad range of plasma states that an experiment is designed to access?

 An important question is when a small change is to be made in $\vec{P}$ how large is the maximum change in $\vec{I}$ that may be required. There is a fundamental control problem when the relative change in the current $||\delta\vec{I}||/||\vec{I}||$ is enormous when compared to the desired change in the properties $||\delta\vec{P}||/||\vec{P}||$.  The maximum possible fractional change in the currents relative to a fractional change in the desired properties is 
\begin{eqnarray}
&&Max\left\{\frac{\frac{||\delta\vec{I}||}{||\vec{I}||}}{\frac{||\delta\vec{P}||}{||\vec{P}||}}\right\} = Max\left\{\frac{||\tensor{C}^{-1}\cdot\delta\vec{P}||}{||\delta\vec{P}||} \frac{||\tensor{C}\cdot\vec{I}||}{||\vec{I}||}\right\}  \nonumber\\
&&\hspace{0.5in}\leq Max\left\{||\tensor{C}^{-1}|| \right\} Max  \left\{ ||\tensor{C}|| \right\}.
\end{eqnarray}
This is the ratio of the largest to the smallest singular value in the Singular Value Decomposition of $\tensor{C}$, which is the condition number.

A problem arises when one wishes to control a component of $\vec{P}$ that is associated with a small singular value of $\tensor{C}$, which means a component of $\vec{I}$ must be very large to drive it.  Unless that component of $\vec{I}$ can be made orthogonal to all components of $\vec{I}$ associated with large singular values of $\tensor{C}$, one cannot control the component of $\vec{P}$ that one wants without making a large unwanted change in a different component of $\vec{P}$.  The hypersensitivity of the currents in the coils relative to the plasma parameters that are to be controlled can be reduced by limiting the components in $\vec{P}$ that one is trying to control to just the most important.  The results of Ku and Boozer \cite{Ku-Boozer:2010,Boozer-Ku:control2011} imply that only a few distributions of $\vec{B}_x\cdot\hat{n}$ require careful control.

A clever choice of the control coils can make the actual sensitivity in plasma control smaller, even much smaller, than the condition-number estimate.  Computer simulations can estimate what types of control are practical and how to design the control coils to provide maximal independent control of the most important elements in $\vec{P}$ over a variety of plasma states.  The limitations and the possibilities for control are clarified by the discussion in Appendix \ref{sec:specification} on the specification of external fields. 
   
%%%%%%%%%%%%%%%%%%%%%%%%%%%%%%%%%%%%%%%%%%%%

%%%%%%%%%%%%%%%%%%%%%%%%%%%%%%%%%%%%%%%%%%%%%%%

\section{Configuration possibilities}

 The Ku and Boozer demonstration \cite{Ku-Boozer:2010,Boozer-Ku:control2011} that only a small fraction of the possible linear magnetic perturbations $\delta\vec{B}\cdot\hat{n}$ to optimized stellarator equilibria have an effect on the degree of physics optimization implies that a large number of degrees of freedom remain in the external magnetic field---even after an optimization---that could be used to achieve other goals.  
 
 Such calculations can be carried out without uncertainties for curl-free magnetic fields.  Curl-free magnetic fields provide an important basis for determining attractive configurations since they can be perturbatively taken to configurations with plasma pressure while reoptimizing the external magnetic field.

%%%%%%%%%%%%%%%%%%%%%%%%%%%%%%%%%%%%%%%%%%%%%%%

\subsection{Possible divertor designs \label{divertors}}

 One such goal in the design and control of plasma divertors.  An important computational study would be what possibilities these remaining degrees of freedom allow for divertors. 
 
 There are two basic types of stellarator divertors:  (1) the island divertor as on W7-X \cite{W7-X:divertor} and (2) the non-resonant divertor.   
The presence of a strong bootstrap current is thought to make island divertors, the type of divertor used on W-7X, impractical.  This may be less true in a power plant than in an experimental facility.   A quasi-steady-state power plant could operate in a single plasma state with a known edge transform.  Both the edge transform and surface shape can be controlled by external coils during startup.  The rotational transform can be controlled by poloidal field coils forming a transformer that allows a controlled change in poloidal flux in the central hole of the torus.

Chaos is unimportant for the magnetic field lines in the W7-X island-divertor but is essential for understanding non-resonant divertors.   Magnetic field lines are chaotic when within a region of non-zero volume each field line has lines that are infinitesimally separated from it that have a separation that increases exponentially with distance along the lines.  

The dictionary definition of chaos as complete disorder is misleading in the application of the concept to non-resonant divertors.  Chaotic magnetic field lines can lie in well collimated flux tubes that extend from the plasma edge to the locations on the surrounding walls.  These tubes define a non-resonant divertor, and the locations where they strike the walls are remarkably robust to changes in the plasma configuration \cite{Bader:2023}. 
 
Non-resonant divertors appear naturally in optimized stellarators due to locations of sharp curvature of the magnetic surfaces near the plasma edge, Bader et al \cite{Bader:2018}.  Due to $\vec{\nabla}\cdot\vec{B}=0$, these are also the locations at which the separation between neighboring magnetic surfaces becomes large.  Boozer and Punjabi \cite{Boozer-Punjabi:2018} used these features to derive an analytic magnetic-field-line Hamiltonian, the poloidal magnetic flux as a function to the toroidal magnetic flux and two angles, $\psi_p(\psi,\theta,\varphi)$.  They have used this Hamiltonian in studies of topological features of non-resonant divertors, such as Cantori, turnstiles, and magnetic field line chaos.  

  An important physics issue for non-resonant stellarator divertors is the electric field that is required for quasi-neutrality and the associated transport in chaotic magnetic field regions \cite{Boozer:divertor}.  Unlike the coil issues in curl-free stellarator models, the way to proceed in addressing this issue is far from certain.
  
 The magnetic surfaces of the well-confined plasma are separated from the sea of chaotic field lines that surrounds it by Cantori, which are like irrational magnetic surfaces that have localized holes through which magnetic field lines can pass, with equal fluxes in both directions, and reach the chamber walls.  These holes are caused by nearby unstable fixed points---field lines that exactly close on themselves after making a number of transits of the torus but from which neighboring lines exponentiate.  In helical symmetry, these would be the X-points of islands, but in the absence of symmetry, the field lines in the neighborhood of an X-point become chaotic.
 
 Using a map that models a non-resonant divertor, Boozer and Punjabi  \cite{Boozer-Punjabi:2018} found that the probability that a field line will be lost from a Cantorus in one toroidal transit has the form
 \begin{equation} 
 P_\ell(\psi) = \left(\frac{\psi-\psi_o}{w_\ell}\right)^d,
 \end{equation}
 where $\psi$ is the toroidal flux enclosed by the Cantorus, $\psi_o$ is the flux enclosed by the outermost magnetic surface, $w_\ell$ is a constant, and $d\approx2$.  But, the outward and inward flux tubes through a Cantorus need not be adjacent.  Adjacent versus non adjacent turnstiles can have different $d$'s \cite{Punjabi:2022}.   
 
 The number of toroidal transits required for a field line to be lost can be written as
 \begin{equation}
 N_\ell(\psi) = \left(\frac{w_\ell}{\psi-\psi_o}\right)^d. \label{N-ell}
 \end{equation}

The number of toroidal transits that would be required for plasma to diffuse to the Cantorus at which a field line is lost is $N_\ell(\psi_\ell) = (\psi_\ell - \psi_o)^2/D_p$.  Equating the $N_\ell$ from this equation with that from Equation (\ref{N-ell}) gives
 \begin{eqnarray}
 \psi_\ell -\psi_o &=& \left(D_p w_\ell^d\right)^{\frac{1}{2+d}};\\
 N_\ell &=& \left(\frac{w_\ell^2}{D_p}\right)^{\frac{d}{2+d}};\\
 P_\ell &=& \frac{1}{N_{\ell}}.  \label{P-ell}
 \end{eqnarray}
 $P_\ell$ gives the size of each of the two escaping flux tubes that carries plasma to the walls.  Each has a magnetic flux is $\delta\psi = \psi_\ell/N_{\ell}$. 
 
In a non-resonant divertor the plasma flows to the wall along one or more pairs of flux tubes, which appear to be very narrow.  The compactness of the flux tubes that carry the plasma to the divertor may seem surprising.  The same narrowness is important in the runaway electron problem in post-disruption tokamaks.  Relativistic electrons from a central region of chaotic field lines exit the plasma along narrow flux tubes when the field evolves slowly compared to the time required for a relativistic electron to cover the chaotic region but exits the plasma over a broad region when the evolution is fast \cite{runaway:2016}.  Both effects are seen in tokamak experiments.

A way to study the collimation of the narrow flux tubes carrying plasma to the divertor is to cover the surrounding chamber wall with a grid.  In almost all grid cells, the normal magnetic field to the wall will be non-zero, and the field lines can be followed from the location where they enter the chamber to the place where they exit after $N_t$ toroidal transits.  Equation (\ref{P-ell}) says the probability of a field line staying in as $N_t\rightarrow\infty$ is zero.  Nonetheless there can be pairs of points on the chamber wall where field lines make $N_t\rightarrow\infty$ transits.  These lines with $N_t\rightarrow\infty$ come extremely close to a magnetic surface, which can be the outermost magnetic surface of the confining plasma and are the lines to be used in divertors.  Another possibility is that lines with $N_t\rightarrow\infty$ come extremely close  to the outermost surface of a magnetic island that lies outside the plasma. 

 In a stellarator the helical field $B_h$ is much stronger than the net poloidal field $B_\iota\propto (B_h/B)^2B$ that determines the rotational transform \cite{Spitzer:1958}.  In an unpublished study, Alkesh Punjabi finds that in most grid cells field lines quickly exit, $N_t\sim1$.  Although the magnetic field lines are in principle chaotic in the region outside those bounded by magnetic surfaces, once a magnetic field line escapes from the Cantori that surround these surfaces, the line strikes the wall within a few toroidal transits, so the exponential separation of neighboring lines does not occur over a sufficient length to destroy the compactness of the exiting flux tubes.

 Understanding how the location of the strike points and the width of the diverted flux tubes are controlled by the freedom in the external coils remains to be established.  Studies of curl-free stellarator configurations would clarify how the location of the $N_t\rightarrow\infty$ points and the widths the holes in the Cantori, the $w_\ell$, are determined.  The control of both is very important for the design of attractive non-resonant divertors.  Although the location of the $N_t\rightarrow\infty$ points is remakably robust, baffles to keep neutrals close to divertor pumps can be better designed the more accurately the location of these points can be defined and the behavior of field lines near these points can be determined, which is related to $w_\ell$.

%%%%%%%%%%%%%%%%%%%%%%%%%%%%%%%%%%%

\subsection{A potentially advantageous stellarator configuration}

%%%%%%%%%%%%%%%%%%%%%%
\begin{figure}
\centerline{ \includegraphics[width=2.0 in]{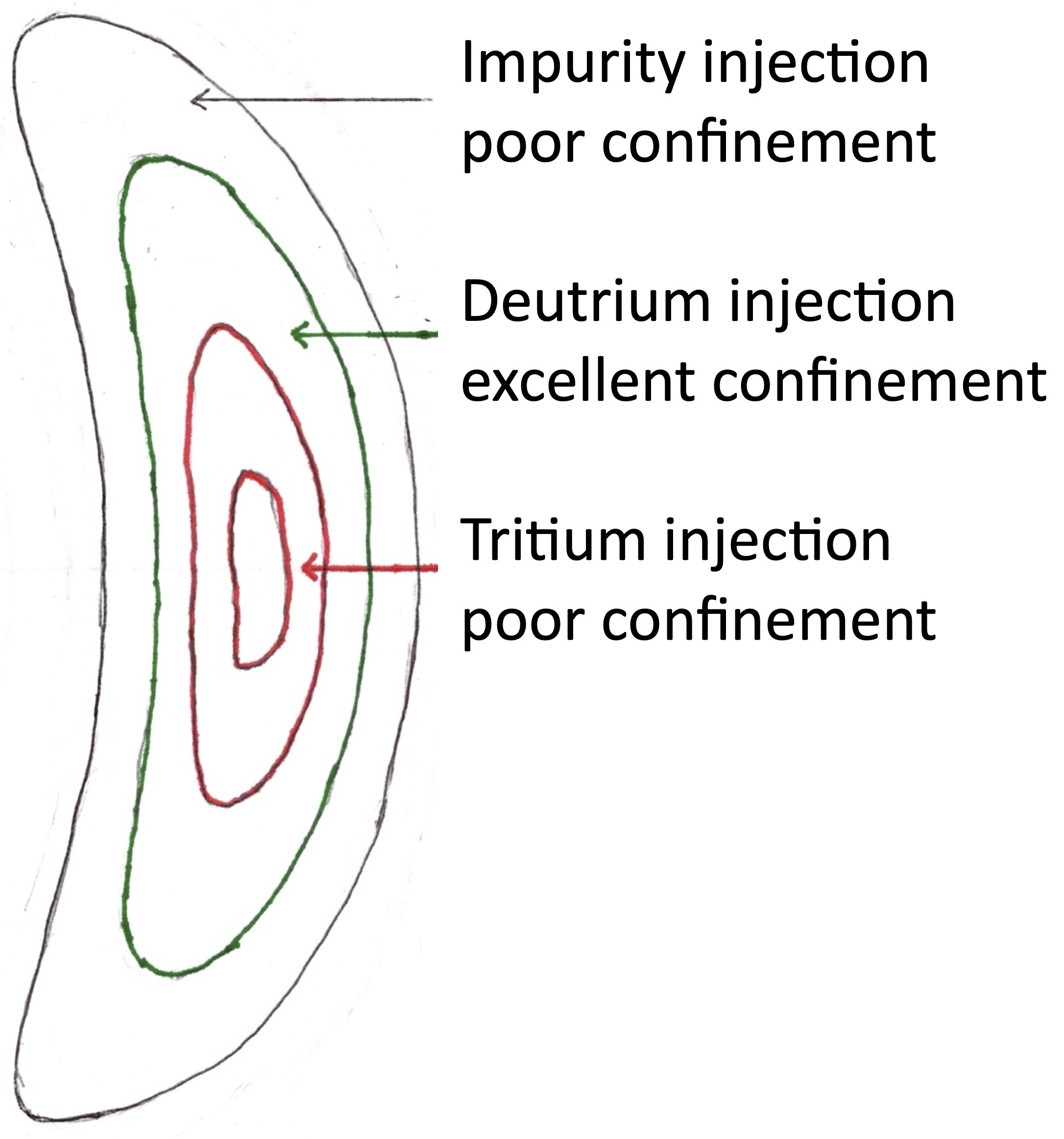}}
\caption{A stellarator plasma could be designed to have three confinement zones.  The innermost zone would have rapid transport and be the target for injected tritium pellets.  The intermediate zone would have sufficiently low transport for a fusion burn and deuterium pellets would be injected in the outer part of this zone.  The outermost zone would have rapid transport and extend into the region of open lines that form the divertor and be the target for injected impurities. } 
\label{fig:zoned}
\end{figure}
%%%%%%%%%%%% 

A stellarator configuration \cite{Boozer:zoned-stell}, which is illustrated in Figure (\ref{fig:zoned}), has been suggested as way to  achieve three goals: (1) Enhance the confinement time of tritium. (2) Have a sufficient density of high-Z impurities to radiate the thermal power escaping from the core as is needed for an acceptable divertor solution while having an extremely low impurity density in the core. (3) Maintain a large fraction of the plasma in a burning plasma state with an optimal tritium fraction.

 When the tritium density equals that of deuterium, the fraction of the tritium that is burnt in an energy confinement time is a fraction of a percent.  This follows from the enormous ratio of the energy of the fusion-produced alpha particles to the temperature of the plasma. The tritium must make many passes through the plasma before it is burnt.  This worsens a major issue for  fusion power, the achievement of tritium self-sufficiency of deuterium-tritium power plants \cite{tritium,tritium2}.   Concepts for increasing the number of tritiums produced per neutron would have a major impact on the feasibility of fusion.
 
  There are two ways to increase the tritium burn fraction: (1) lower the tritium fraction in a deuterium-tritium plasma, at the cost of requiring better energy confinement to achieve a self-sustained burn or (2) increase the confinement time of the tritium.  Both are discussed in \cite{Boozer:zoned-stell}, but the zoned concept illustrated in Figure (\ref{fig:zoned}) is designed to increase the tritium confinement time.
  
  In standard stellarator reactor designs, the helium and high-Z impurities are flushed out of the plasma by a 50-50 deuterium-tritium mixture.  Tritium is many orders of magnitude more expensive than deuterium.  The zoned concept tries to avoid the bizarre use of tritium as a flushing fluid.  
 
 Before the attractiveness of the zoned concept can be assessed two physics issues must be better understood:  (1) the physics of the interdiffusion of different species and (2) the localization and the depth of plasma penetration by pellets.

%%%%%%%%%%%%%%%%%%%%%%%%%%%%%%%%%%%%%%%%%%%%%%%%%

\subsection{Other issues that should be studied}

The consistency of various features of plasma configurations is important to study.  An example, is it possible to have a large shear in the rotational transform in a curl-free configuration with small neoclassical transport?  Another is the tendency of optimized stellarators to have outer surfaces with high surface curvature.  These regions are important for the formation of non-resonant divertors.

Reference \cite{Boozer:omnigeneity} discussed optimal ion and electron temperatures and their control together with the generality of gyroBohm scaling.  These issues have significant impacts on stellarator design and are active areas of research, but the level of certainty that can be achieved is far below that in coil design for curl-free stellarator configurations.  

A related issue that should be studied is whether stellarators can be designed so Alfv\'enic modes can be driven by the energetic alpha particles but only in the central region.  Central Alfv\'enic modes do not directly cause an enhancement of the energetic alpha losses to the walls.  To the extent these alpha-driven Alfv\'en waves are damped by the viscosity associated with background deuterium and tritium ions, direct ion heating is achieved.  Normally, the alpha energy is transferred to the electrons, which indirectly transfer it to the ions.  This indirect method fails to heat ions when the electron-ion equilibration time is long compared to the ion energy confinement time, which can be the case when the electrons become too hot \cite{Boozer:omnigeneity}.

%%%%%%%%%%%%%%%%%%%%%%%%%%%%%%%%%%%%%%%%%%%%%%%%%%%%

\section{Effects of sudden plasma termination}

Sudden plasma terminations, disruptions, are a potential fatal flaw of tokamak power plants.  This issue can to a large extent be circumvented by the use of stellarators.  Nevertheless, there are potential disruption issues for stellarators that should be studied computationally.

The plasma temperature in a stellarator can suddenly collapse for two reasons.  (1) A piece of wall material could fall into the plasma causing a radiative collapse with magnetic surfaces remaining intact.  (2) An ideal instability that does not self-saturate can sufficiently contort the magnetic surfaces that they resistively reconnect on a timescale set by the growth rate of the ideal instability \cite{Boozer:surf-loss:2022,Jardin:fast breakup:2022}.  

Unlike the vertical field of a tokamak, the helical field in a stellarator exerts a force that centers the plasma in the chamber.  The sudden loss of the plasma temperature produces far less dramatic effects than in a tokamak.  Nevertheless, computations should study the effect of the sudden change in the diamagnetic toroidal flux due to the loss of the plasma pressure as well as the forces and power deposition on the chamber walls.

%=========================

\appendix

%%%%%%%%%%%%%%%%%%%%%%%%%%%%%%%%%%%%%%%

\section{Freedom in external magnetic fields \label{sec:B-freedom}}

Two time-independent equations, $\vec{\nabla}\cdot\vec{B}=0$ and $\vec{\nabla}\times\vec{B}=\mu_0 \vec{j}$, define the full freedom of magnetic fields that can be produced in a plasma chamber by external coils.  Within the region enclosed by a toroidal surface that lies closer to the plasma than any coil, the coil-produced external magnetic field must have the form $\vec{B}_x=\vec{\nabla}\phi_\ell$, where $\phi_\ell$ is a solution to Laplace's equation, $\nabla^2\phi_\ell=0$.  

The general solution $\phi_\ell$ in a toroidal region is the sum of a single-valued term $\phi$ and two multivalued terms, which depend linearly on $\theta$ or $\varphi$.  One multivalued term increases by $\mu_0 G_0$ each toroidal circuit of the torus and the other by $\mu_0 I_0$ each poloidal circuit.  A non-zero $G_0$, which is the net current coming through the hole in the center of the toroidal enclosing surface, determines the toroidal flux within that surface.  A non-zero $I_0$, which is the net toroidal current outside the enclosing surface, determines the magnetic flux through the hole in the center of the enclosing surface.   The effect of $I_0$ within the enclosing surface can represented by the single-valued potential $\phi$.

Within an additive constant, $\phi$ is unique when $\hat{n}\cdot\vec{\nabla}\phi$ is given on an enclosing boundary with $\hat{n}$ the unit normal to that boundary.  This is called a Neumann boundary condition.

The theory of Laplace's equation implies that there is far more freedom in coil design than has been exploited in stellarator design.  The general externally produced magnetic field has the form
\begin{eqnarray}
\vec{B}_x(\vec{x}) &=& \vec{B}_b + \vec{B}_d,
\end{eqnarray}
where $\vec{B}_b$ is the field produced by a basic coil set, such as simple toroidal field, general modular, or helical coils.     The magnetic field $\vec{B}_d$ is a magnetic field that can be produced by magnetic dipoles normal to the enclosing surface.  These dipoles control the normal magnetic field on the surface.  Giving a distribution of dipoles is equivalent to a Neumann boundary condition.  

%%%%%%%

\subsection{The basic field $\vec{B}_b$} 

The field $\vec{B}_b$ must produce the toroidal magnetic flux.  Although the basic field can be produced by any specified external coil set, a simple analytic field can be used for general studies.  This field is based on ordinary cylindrical coordinates  $(R,\varphi,Z)$, where $R=0$ is in the central hole of the torus and $G_0$ is constant, the number of Amperes of current coming up through the central hole of the torus:
\begin{eqnarray}
&& \vec{B}_b = \vec{\nabla}\phi_b;\\
&& \phi_b = \frac{\mu_0 G_0}{2\pi}\Big\{\varphi \\
&& \hspace{0.1in}+ \left( \left(\frac{R_{in}}{R}\right)^N + \left(\frac{R}{R_{out}}\right)^N\right)\frac{ \sin N\varphi}{N} \Big\}, \hspace{0.3in} \label{rippled-B}
\end{eqnarray}
where $\vec{\nabla}\varphi = \hat{\varphi}/R$.  The locations of the $N$ toroidal field coils are centered on the places where $\cos N\varphi =1$ and the spaces between the coils are centered on places where $\cos N\varphi=-1$.  The inner legs of the toroidal field coils at $R=R_{in}$ and the outer legs are at $R=R_{out}$.

%%%%%%%%%%

\subsection{The dipole field $\vec{B}_d$}

A general specification of the dipole field is given by dividing the enclosing toroidal surface into $L$ cells with the center of the $\ell^{th}$ cell located at the point $\vec{x}_\ell$ where the unit normal to the surface is $\hat{n}_\ell$.  As $L\rightarrow\infty$ the field due to dipoles normal to the enclosing surface is
\begin{eqnarray}
\vec{B}_d(\vec{x}) &=& \frac{\mu_0}{4\pi}  \sum_{\ell=1}^L  \frac{3 (\hat{n}_\ell\cdot\hat{\Delta}_\ell)\hat{\Delta}_\ell  - \hat{n}_\ell}{\Delta_\ell^3} m_\ell ; \\
\vec{\Delta}_\ell(\vec{x}) &\equiv& \vec{x} - \vec{x}_\ell
\end{eqnarray}
where $m_\ell\hat{n}_\ell$ is the dipole moment of the $\ell^{th}$ cell.

The dipole moment of grid cell $\ell$ is given by an integral over the area of the cell, Appendix \ref{dipole},
\begin{equation} 
\vec{m}_\ell = \int \kappa_s d\vec{a}_\ell,
\end{equation}
where $\kappa_s(\theta,\varphi)$ is a periodic function of the poloidal and toroidal angles and is called the single-valued current potential.  The single-valued current potential is equivalent to a surface current on the enclosing surface
\begin{equation} 
\vec{J}_s = \vec{\nabla}\kappa_s\times\hat{n},
\end{equation}
where $\vec{J}_s$ is the integral of the current density across the enclosing surface when treated as a thin shell.  A sufficiently fine grid is required to accurately represent the surface current.  The value of the single-valued current potential at the center of the $\ell^{th}$ cell can be taken to be $\kappa_\ell \equiv |\vec{m}_\ell|/a_\ell$, where $a_\ell$ is the area of the $\ell^{th}$ cell. 

When the single-valued current potential $\kappa_s(\theta,\varphi)$ can be represented as a Fourier series, as is conventionally assumed \cite{Merkel:1987}, the accuracy of the surface current is ensured.  But as illustrated by the well-known Gibbs phenomenon, Fourier series are an extremely poor way to represent a function that has regions in which it is constant.  Consequently, the Fourier representation essentially eliminates the possibility of having currents in localized regions of the bounding surface despite localized currents being a natural way to produce very attractive coils.  The specification of the dipole field by dipoles on a grid eliminates this artificial constraint and is called a specification by current potential patches in Todd Elder's thesis \cite{Elder:thesis2023}. 

%%%%%%%%%%%%%%

\subsection{Specification of external magnetic field \label{sec:specification}}

The external magnetic field has two roles: (1) The field from the basic coil set $\vec{B}_p$ must provide the toroidal magnetic flux.  For that, at least one coil must encircle the plasma poloidally.  (2) The dipole field must supply whatever normal magnetic field is required on the desired plasma surface to ensure that surface is a magnetic surface.  Other than providing the toroidal flux, the coils producing $\vec{B}_p$ are unconstrained.  

The dipolar field must produce a specified normal magnetic field distribution on the plasma surface.  The dipoles can be specified by a matrix vector $\vec{d}$, which has one component for each of the $L$ dipoles.  The normal component of the dipolar field on the plasma surface can be specified by a matrix vector $\vec{b}$, which retains $K$ components of the Fourier decomposition of the required normal field.  Fourier series of analytic functions converge exponentially, so the accuracy with which the normal field is specified generally increases exponentially with $K$, so $K$ need not be extremely large.

A $K$ by $L$ matrix $\tensor{M}$ can be calculated in a straight-forward manner  that relates the matrix vector $\vec{d}$, which has $L$ components, one for each dipole, and a matrix vector $\vec{b}$ that has $K$ components, one for each of the retained Fourier modes.  Once the matrix $\tensor{M}$ is determined, it can decomposed using a Singular Value Decomposition (SVD):
\begin{eqnarray}
&& \tensor{M}\cdot\vec{d} = \vec{b} \\
&& \tensor{M} = \tensor{U}\cdot \tensor{S} \cdot \tensor{V}^\dag.
\end{eqnarray}
The matrix $\tensor{S}$ has only positive diagonal entries, called singular values, of which no more than $K$ can be non-zero. The matrices $\tensor{U}$ and $\tensor{V}$ are unitary and the superscript $\dag$ means a transpose, so $\tensor{U}\cdot\tensor{U}^\dag=\tensor{1}$.  %The ratio of the largest to the smallest non-zero singular value is called the condition number $N_c$. 

The unitary property of the $\tensor{U}$ and $\tensor{V}$ allow the definition of the left and right singular vectors that are associated with the singular values of $\tensor{S}$: 
\begin{eqnarray}
\vec{d}_s &\equiv& \tensor{V}^\dag\cdot\vec{d} \hspace{0.2in} \mbox{and  }\\
\vec{b}_s &\equiv& \tensor{U}^\dag \cdot \vec{b} \hspace{0.2in} \mbox{with }\\
\vec{d}_s &=& \tensor{S}^{-1}\cdot \vec{b}_s.
\end{eqnarray}
The larger the singular value the easier it is for the associated distributions of dipoles to produce the associated magnetic field distribution.  The singular values generally decrease exponentially---only a moderate number of the singular values are sufficiently large for a practical coil set.

The smallest singular value that must be retained to adequately fit a given plasma configuration determines how close the coils must be to the plasma and their complexity.  This is the issue of the efficiency of a coil set \cite{Boozer:RMP,Control:2010,Boozer:design,Landreman:eff-B}.

When the entire bounding surface is covered with the grid of dipoles, then a solution will be obtained with the smallest singular value as large as possible with a given coil-plasma separation.   As discussed by Todd Elder \cite{Elder-Boozer:2024}, some fraction $f_d$ of the dipoles can be eliminated from the $\tensor{M}$, the ones with the smallest magnitudes or the ones most inconveniently located, and a new SVD calculated.  As long as the smallest singular value remains sufficiently large, the remaining patches of dipoles can accurately produce the normal magnetic field on the plasma surface to the required accuracy.

To further reduce either the required coverage of the bounding surface with dipoles or its maximal separation from the plasma, the plasma configuration should be changed to reduce or eliminate its dependence on a field distribution associated with the smallest singular value.

The fineness of the required dipole grid is largely determined by the number of magnetic field distributions are associated with sufficiently large singular values to be of practical importance.  The magnetic field itself is accurately represented by a relatively coarse grid.  The grid is defined by the length  $\delta_\ell$ of a side of a grid cell $\ell$.  Following a 2011 discussion by Boozer and Ku \cite{Boozer-Ku:control2011}, the approximate answer is $\delta_\ell  \lesssim \Delta^{(p)}_\ell/1.3$, where  $\Delta^{(p)}_\ell$ is the distance cell $\ell$ is from the plasma surface, on which the external magnetic field is to be controlled.  A finer grid may be needed to calculate the required surface currents on the bounding surface with the desired accuracy

The magnetic energy $\int (B^2/2\mu_0)d^3x$ and the forces on the coils can be relatively easily calculated as discussed in Section 2.3 of \cite{Boozer:wall-interaction2021}.\\

\subsection{Derivation of the dipole moment of the current potential in a cell} \label{dipole}

The single-valued part of the current potential $\kappa_s$ in a small region on a surface can be represented by a magnetic dipole.  Let $\hat{z}$ be the normal to the surface and use cylindrical coordinates.  The magnetic moment of a current density is
\begin{eqnarray}
\vec{m} &=& \frac{1}{2} \int r \hat{r}\times\vec{j} d^3x, \mbox{   and   } \\
\vec{j} &=& (\vec{\nabla}\kappa \times \hat{z})\delta(z), \mbox{   so   }\\
\vec{m} &=&- \frac{\hat{z}}{2} \int r \frac{\partial \kappa(r,\theta)}{\partial r}\delta(z) rdrd\theta dz\\
&=&-\frac{\hat{z}}{2} \int r^2 \left(\int_0^{2\pi} \frac{\partial \kappa(r,\theta)}{\partial r}d\theta\right) dr\\
&=&-\frac{\hat{z}}{2} \int 2\pi r^2 \bar{\kappa}(r) dr\\
&=& -\pi \hat{z} \int \Big(\frac{d}{dr}( r^2 \bar{\kappa})- 2r\bar{\kappa} dr\\
&=& \hat{z} \int \kappa(r,\theta) r dr d\theta,
\end{eqnarray}\\
where $2\pi\bar{\kappa}\equiv \int_0^{2\pi}\kappa(r,\theta)d\theta$.  Since $(\hat{r}\times\hat{\theta})\cdot\hat{z}=1$, $\hat{r}\times(\vec{\nabla}\kappa\times\hat{z}) =-\hat{z} \partial\kappa/\partial r$.  The current potential $\kappa(r,\theta)$ is assumed to be zero outside of the region of the radial integration.  The area element for this patch on the surface is $da=rdrd\theta$.

%===============================================
%%%%%%%%%%%%%%%%%%%%%%%%%%%%%%%%%%%%%%%%%%%%%%%%%%%%%%%
\section*{Acknowledgements}

This material is based upon work supported by the grant 601958 within the Simons Foundation collaboration \emph{Hidden Symmetries and Fusion Energy} and by the U.S. Department of Energy, Office of Science, Office of Fusion Energy Sciences under Awards DE-FG02-95ER54333 and DE-FG02-03ER54696 and Sub-Awards Princeton University PU SUB0000764 and Princeton Plasma Physics Laboratory PPPL P-240000750. 

%%%%%%%%%%%%%%%%%%%%%%%%%%%%%%%%%

\end{document}